# Astro2020 Science White Paper

# Comparing key compositional indicators in Jupiter with those in extra-solar giant planets.

**Thematic Areas:**  ☒ Planetary Systems   ☒ Star and Planet Formation

☐ Formation and Evolution of Compact Objects   ☐ Cosmology and Fundamental Physics

☐ Stars and Stellar Evolution   ☐ Resolved Stellar Populations and their Environments

☐ Galaxy Evolution   ☐ Multi-Messenger Astronomy and Astrophysics

**Principal Author:**
Name: Jonathan I. Lunine
Institution: Cornell University
Email: jlunine@astro.cornell.edu
Phone: 607-319-0439

**Co-authors:** (names and institutions)

Tom Greene, NASA Ames Research Center
Charles Beichman, California Institute of Technology
Jacob Bean, University of Chicago
Heidi B. Hammel, AURA
Mark S. Marley, NASA Ames Research Center

**Abstract**

Spectroscopic transiting observations of the atmospheres of hot Jupiters around other stars—first with Hubble Space Telescope and then Spitzer—opened the door to compositional studies of exoplanets. The James Webb Space Telescope will provide such a profound improvement in signal-to-noise ratio that it will enable detailed analysis of molecular abundances, including but not limited to determining abundances of all the major carbon- and oxygen-bearing species in hot Jupiter atmospheres. This will allow determination of the carbon-to-oxygen ratio, an essential number for planet formation models and a motivating goal of the Juno mission currently around Jupiter.

**Introduction**

Just a half-century ago, spectroscopic observations of the solar system's giant planets (Gillett et al. 1969) were roughly equivalent to today's transiting spectroscopic observations of giant planets around other stars (Beichman et al. 2014). By the 1990s, the science of giant planet atmospheres had moved to direct sampling: the Galileo entry probe descended through Jupiter's atmosphere and its mass spectrometer measured a range of molecular species and noble gases. However, the probe measured such a low abundance of water (Wong et al. 2004) that it was thought to represent an anomaly caused by atmospheric dynamics (Figure 1). Water is of keen interest because it is the primary molecular carrier of oxygen in the observable part of Jupiter's envelope. Oxygen is of course the third most abundant element in the universe and is thought to be a key to the formation of giant planets under the core accretion paradigm. Currently, both ground-based high-resolution infrared spectroscopy (Bjoraker et al. 2018) and close-up microwave measurements by NASA's Juno spacecraft (Janssen 2016) are being applied to determine Jupiter's water abundance.

Against the backdrop of these extensive efforts to complete the inventory of major elements in Jupiter by measuring the planet's water abundance, JWST is poised to open up a new era of highly sensitive transit spectra of extrasolar giant planets. Despite their vastly greater distance from us, extrasolar giant planets in close orbits around their parent stars (hot Jupiters) present a vastly simpler case for determining the abundance of water because their atmospheric temperatures are well above the condensation point for cloud formation. Indeed, a nearly complete inventory of carbon- and nitrogen-bearing species may be obtained, as has been known since the first transit studies. Companion white papers submitted cover the technique of transit observations (Greene et al.) and alternative techniques of direct spectroscopy for giant planets in more distant orbits (Beichmann et al.). In this white paper we focus on the scientific benefits of such observations, particularly in the context of comparison with the results from Jupiter.

Guaranteed Time Observation (GTO) programs are listed at https://jwst.stsci.edu/observing-programs/approved-gto-programs. Programs that potentially or explicitly address the science described here include 1177, 1185, 1195, 1201, 1224, 1246 (Jupiter) 1274, 1277, 1278, 1280, 1281, 1292, and 1353.

**Formation of planets and C/O ratios**
The ratio of carbon to oxygen in a gaseous condensing system, such as a protoplanetary disk, is important because it determines the abundances of solid-forming material (Figure 2). However, in a closed system, such as might comprise a molecular cloud clump and protoplanetary disk, the system-wide C/O ratio remains constant and so partitioning would be evident only due to spatial variations and enrichment (or depletion) in the condensate vs gaseous phase. These effects would themselves vary in time over the life of the disk.

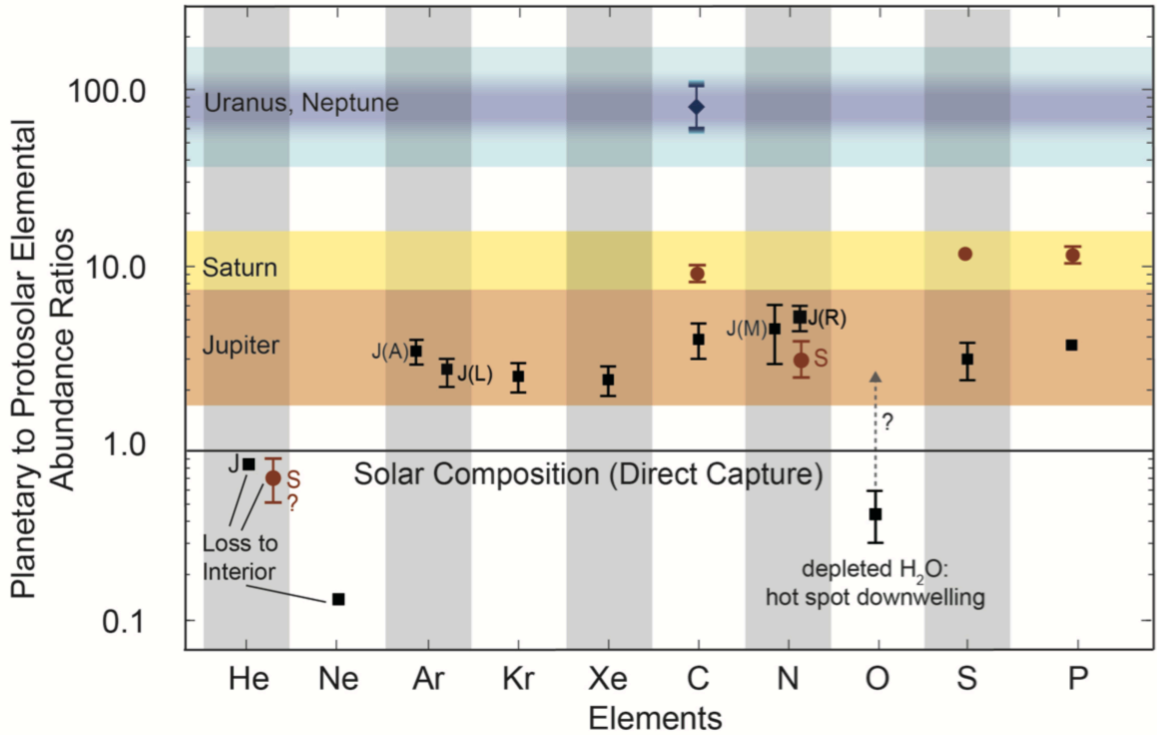

**Figure 1. Major element abundances in the giant planets.** Abundances are relative to the accepted primordial solar values as discussed in the source of this figure (Atreya et al. 2019), i.e., an element that is precisely at the primordial solar value will be at unity. Horizontal colored bars and vertical striping are present just to guide the eye. Carbon is from methane, nitrogen from ammonia, oxygen from water, sulfur from hydrogen sulfide, and phosphorus from phosphine. Suggested processes that deplete He and O are labelled. Two Ar values are shown for different assumed Ar/H in the Sun. Nitrogen in Jupiter is derived from ammonia via the Galileo probe mass spectrometer "J(M)" and attenuation of the probe radio signal "J(R)".

Processes that act to change C/O in a disk include condensation of ices at multiple nebular snowlines, for example those of water and carbon monoxide. Because these snowlines are at widely separated radial distances in the disk, transport timescales differ in the respective locations. As a result, the C/O value in the gas will vary over time and space (Oberg et al., 2011; Ali-Dib et al., 2014). As a consequence, bodies such as giant planets formed from the gas may exhibit different C/O values from the bulk disk value. Because the molecular composition of a giant planet represents a gross re-equilibration from the molecular composition of the disk, the latter is not preserved—but the elemental ratios are.

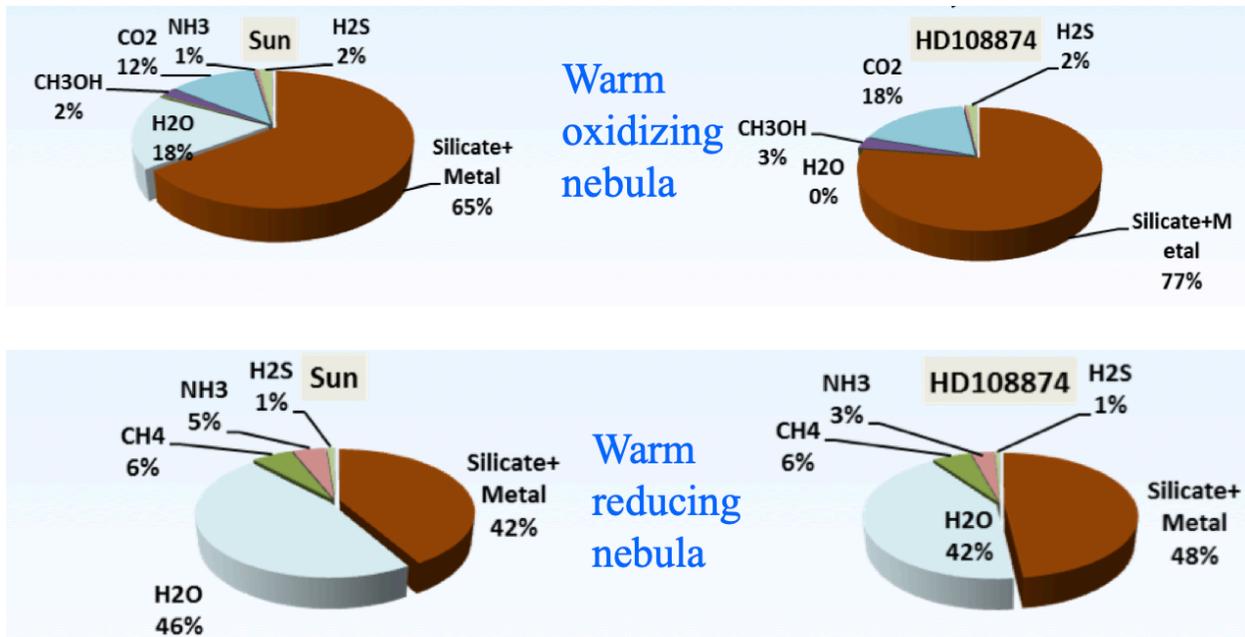

**Figure 2: Percentages of various condensates in a protoplanetary disk with C/O =0.55 (left panels) and C/O = 0.71 (right panels).** The oxidizing nebula is chemically quenched so that the primary carbon-bearing molecule is CO; in the reducing nebula it is $CH_4$. As a consequence, the water abundance in the disk represented by the upper panels is much more sensitive to C/O. Based on calculations reported in Johnson et al. (2012).

Variations in the gas-phase value of C/O—whether through snowline-driven evolution or other processes—may be expressed directly through gas accretion onto giant planets or indirectly through the corresponding effect on condensates. The challenge in interpretation is that, as a single ratio, C/O cannot diagnose the range of temporally and spatially varying processes that might change it (Figure 3). However, as one diagnostic among several compositional indicators (including the mass of a giant planet core and the heavy element abundance in the envelope) C/O can be useful (Helled and Lunine, 2014).

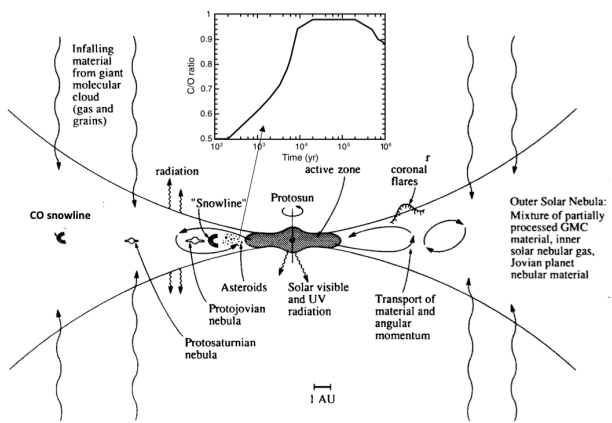

**Figure 3. Schematic of the solar system's protoplanetary stage, with various disk processes shown, modified from Lunine (1989).** Included are notional positions of the water ice and CO-ice snowlines, relative to Jupiter and Saturn. As a consequence of the presence of these snowlines, C/O at 2 AU evolves with time, as shown in the inset from Ali-Dib et al. 2014).

# C/O in Jupiter and extrasolar giant planets

The Juno mission currently in orbit around Jupiter has constrained the mass of Jupiter's core and the envelope heavy element abundance, and is on its way to constraining the water abundance as well. Although ambiguities will remain in these determinations, they represent the most detailed compositional information we have on any giant planet. Of keen interest is whether the water abundance will imply a nonsolar C/O value or not; to address this will require fitting the entire suite of abundances of noble gases and presumed nebular molecular carriers of C, O, P, S, and N (Figure 1) to planetesimal formation and ice trapping models (e.g., Mousis et al. 2012).

With respect to extrasolar giant planets, the claimed presence of differences in C/O between their atmospheres and their parent stars is controversial, in large part because the quality of current transit observations makes it difficult to interpret the spectra of all but a handful of such objects (Madhusudhan et al., 2016). JWST will, however, be able to constrain the abundances of $CH_4$, CO, $CO_2$, $H_2O$ and $NH_3$ and other species well, especially if a broad enough wavelength range is used to overcome the effects of clouds (Greene et al. 2016). It may also be possible to detect spectral signatures and determine the abundances of $H_2S$ and $PH_3$ with JWST (Figure 4).

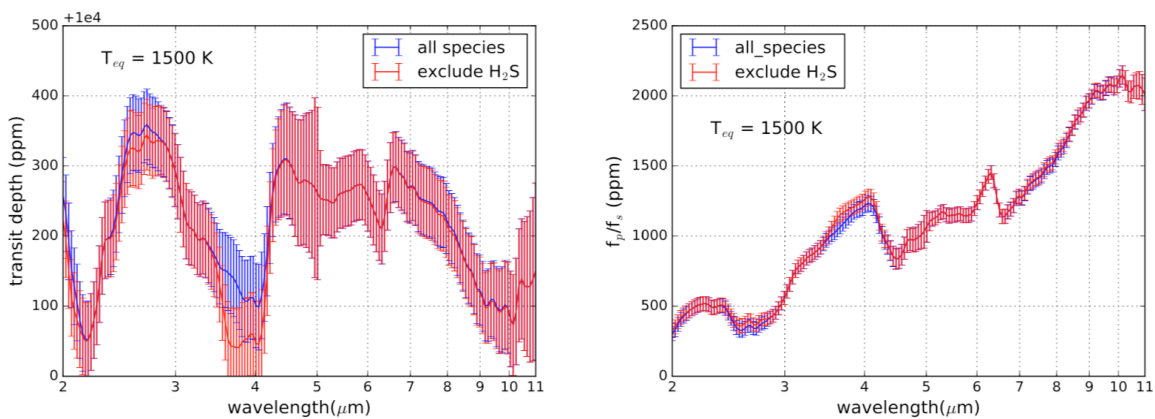

**Figure 4. Synthetic spectra of hot Jupiter transits.** These spectra have spectral resolution 100 and simulated JWST noise, for primary (left) and secondary (right) transits of a hot Jupiter with an equilibrium temperature of 1500 K. The blue curve includes all species considered in the analysis, published as Wang et al. (2017); these are $CH_4$, CO, $CO_2$, $H_2O$, HCN, $H_2S$, $N_2$, $NH_3$ and $PH_3$; the red curve excludes $H_2S$.

The wealth of spectral data in the atmospheres of extrasolar giant planets that awaits unlocking by JWST observations presents an opportunity to determine the basic mechanisms of giant planet formation across multiple planetary systems. One key question that JWST can definitively answer is whether indeed C/O is different in giant planets than in their parent stars; specifically, whether C/O is significantly greater than 0.55, which is the value for sun-like stars (Madhusudhan et al., 2016).

Although any individual giant planet may not present enough spectral indicators to provide unique tests of complex and numerous protoplanetary disk processes, systematics across many such planets around different stars can indicate certain processes as of key importance.

Madhusudhan et al. (2014) argued that migration of Jupiters through the protoplanetary disk will leave key indicators in place that can be found in their spectra if such planets end up as hot Jupiters. These include abundances of heavy elements that exceed that of their parent star, and C/O close to that of the parent star (Madhusudhan et al 2016); these systematics are apparent on a diagram of C/H vs O/H (Figure 5).

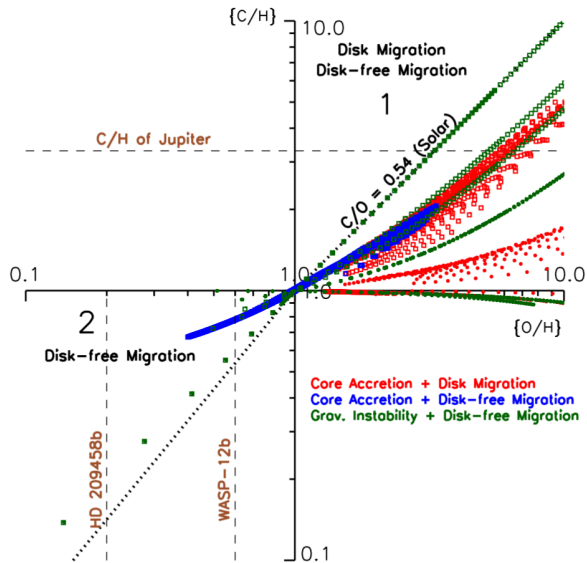

Figure 5. Systematics in oxygen and carbon in giant planet atmospheres for different formation and migration scenarios. Model results are colored in red, blue and green; constraints for several objects are labelled in brown. , Figure from Madhusudhan et al. (2014; 2016).

Almost certainly the aggregate of observed exoplanet atmospheric O and C abundances will not neatly fall on the predicted lines, because other processes besides migration may affect the heavy element abundances (e.g., those discussed above associated with snowlines). Nevertheless, any comprehensive set of results for a sufficiently large number of extrasolar giant planets will be informative. Indeed, the graph as shown provides an interesting prediction for Juno's eventual determination of O in Jupiter.

**Summary**

Spectral observations of extrasolar giant planets provide a hint of interesting results to come, when JWST is launched and provides spectra of high signal-to-noise, resolution and wavelength range. Various processes in protoplanetary disks will affect the abundances of the major elements in these atmospheres, the indicators of which will be in the spectra. While C/O is itself a key diagnostic, taken alone it cannot provide unique information on the wealth of such processes, but is instead is determined ambiguously by the sum total of a giant planet's formation and early dynamic history in the disk. However, observations of many giant planets in many systems may produce systematics that allow the relative importance of the various processes to be addressed. Juno's uniquely detailed information on Jupiter's interior structure, core size, aggregate heavy element abundance and (eventually) O abundance provide the ground truth for understanding data from extrasolar giant planets. This not only means spectral data, but for bodies with both mass and radius determined, the bulk density as well. Together, Juno and JWST have the potential to open up a new era in which giant planet systematics will inform formation and evolution models in somewhat the same way that astronomical observations of large numbers of galaxies have informed models of their formation and evolution.

planets: Detectability of $H_2S$ and $PH_3$ with the James Webb Space Telescope. Astrophys. J. 850: 199 (15pp), 2017.

Wong, M. H., Mahaffy, P.R., Atreya, S.K., et al. Updated Galileo probe mass spectrometer measurements of carbon, oxygen, nitrogen, and sulfur on Jupiter. Icarus, 171, 153-170, 2004.